# A Survey: Embedded Systems Supporting By Different Operating Systems

Qamar Jabeen, Fazlullah Khan, Muhammad Tahir, Shahzad Khan, Syed Roohullah Jan

Department of Computer Science, Abdul Wali Khan University Mardan

Zaara1890@yahoo.com, fazlullah@awkum.edu.pk

---

**Abstract:** In these days embedded system have an important role in different Fields and applications like Network embedded system, Real-time embedded systems which supports the mission-critical domains, mostly having the time constraints, Stand-alone systems which includes the network router etc. A great deployment in the processors made for completing the demanding needs of the users. There is also a large-scale deployment occurs in sensor networks for providing the advance facilities, for handled such type of embedded systems a specific operating system must provide. This paper presents some software infrastructures that have the ability of supporting such types of embedded systems.

## 1. Introduction:

Embedded system are computer systems designed for specific purpose, to increase functionality and reliability for achieving a specific task, like general purpose computer system it does not use for multiple tasks. Now days there are many types of distributed embedded system increased the performance and usability. A system is embedded not only by hardware it have also embedded software.

There are a variety of embedded systems used in industrial, commercial areas. e.g Mobile Phones and different type of Network Bridges are embedded used by telecommunication systems for giving better requirements to their users. We use digital cameras, MP3 players, DVD players are the example of embedded consumer electronics. In our daily life its provided us efficiency and flexibility and many features which includes microwave oven, washing machines dishwashers. Embedded system are also used in medical, transportation and also used in wireless sensor network area respectively medical imaging, vital signs, automobile electric vehicles and Wi-Fi modules.

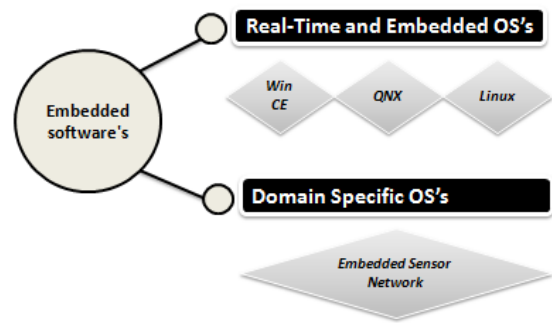

**Figure 1:** *Taxonomy of Embedded Software's*

## 2. Embedded software organization for embedded systems:

Embedded systems needs higher reliability because most of them are safety critical systems therefore consists on microprocessors. Some embedded system are designed from scratch and software specifically developed for those are devices. Different high level programming languages, which fulfils the requirements of that particular software are used by the programmer, having the ability of exception handling and should have the reliable code [1]. Software can be designed using off the shelf components. Operating systems are the example of the reusable software components, which doesn't developed from scratch.

Software should not the platform-dependent it should be the capability of evolvability either it would run on the LAN or shared wide-wide area network. It also should be guaranteed that it provide the same services with respect to the environment of the user, according to their requirement. Accuracy and precision are important when environment changes.[2]

Embedded operating systems are developed for handling the particular application. There for OS used in embedded systems is most different than the general purpose OS used in our desktops computers. And these systems also have the real time properties according to the system requirements. Such types of operating systems are divided into two types, embedded operating system and real time operating systems.

## 2.1 Real time and embedded operating systems:

Embedded systems are mostly used Real-time operating systems, QNX [3], WinCE[4], Vxworks [5] are most widely used embedded operating systems. All embedded devices having operating system which can be supported by that device. An embedded real time device must have the RTOS. Embedded OS having reliability, portability and configurability are the important features in it. RTOS should show the accurate logical results, in specially safety critical systems because in accuracy May loss causes of lost a life like air craft system. The operating systems giving above support the large systems consisting on higher CPU and memory size. And also multitasking, TCP/IP networking, memory protection can be supported by it.

Window NT [6] is not real time operating system, but due to industrial need, window NT should be evolved for achieving some features of real time systems.

Linux operating system also not supported real time system for this purpose RTAI (real time application interface) [7] are designated which is an extension of Linux. Embedded Linux is also an evolution in Linux but it doesn't support real time application like RTAI.

Commercial off-the-shelf components used [8] for achieving the target of distributed real time and embedded systems, such as mission critical dynamic domains which includes the radar processing, online financial trading etc. Clock synchronization [9] is an important feature of distributed real time operating systems. Clock synchronization should be fault tolerated for giving the accuracy. When CPU [10] is highly loaded and unexpected interrupts disturbed time due to this RTOS exceed their limits then more energy will be consumed by it as compared to the normal execution.

In heterogeneous environment embedded systems are more complex in sense of their reliability and security which in more demanding feature in this environment. Over a decade performance of ICT is main focus of researchers. But recently, security and reliability become the higher issue. Some projects proposed by [11] used the micro-kernel for reliability and dependability, [12] proposed Minix OS and [13] suggest the L4 OS for higher reliability.

### 2.1.1 WinCE

November 1996 in handheld PC window CE was introduced. There are many extensions of windows CE. Embedded systems have the restriction of space. Windows CE developed for achieving the given aim. This is a thin client operating system supports priority inheritance and developed for embedded applications like entertainment and communication [14].

Window CE is real-time operating systems which have the functionality of operate the embedded 32 bit processor. Zigbee is an embedded system based on ARM having the serial communication functionality which improves the transmission area and reliability of helicome used as wireless sensor network module. Due to ARM device has of low power consumption (now a days it is an issue of green computing) high performance and lower cost. It also includes the functions data receiving, showing, and the ability of information storage [15].

In zigbee application design algorithms implementation for serial communication and interface design it uses the embedded visual C++. Different languages used by programmer for different algorithms.[16,17]

Systems depending on window CE have 4 layers 1) Hardware layer includes all the hardware used in those systems, such as processors. 2) Operating system is an interface for users. 3) Applications layer consists on the applications which depend upon different types of performances on the device. 4) Hardware support layer consists on interaction between hardware and OS used by that system [18,19].

Basic needs of storing, manipulating, retrieving the data have every operating system. These operations can be performed by replicating the data on a backend server. Enterprise and personal application used such type of database respectively for applications which provided financial services and web-based like entertainment, games etc. For this purpose a relational database like SQL server used by window CE. The constraints on space using by the program is followed by this engine. The architecture of SQL engine includes query optimization, relational storage (transaction and indexing) and backend data replication [20].

Another system used master slave architecture call Computerized Numerical

Control (CNC) using embedded real time operating system.[21]

## 2.1.2 Linux:

Few years ago the use of Linux in embedded device became most popular. People needs accessing their own account anywhere so it is used as a functional OS in networked systems. Linux have an important feature of flexibility, because we can interrupt in kernel at runtime and unload a module. Linux is open source so its code is worldwide available, also portable and scalable. It can run on every processor and also license free.

Differentiability between the hardware make the operating system most sophisticated (mobile devices, severs etc). Different configurators are developed for the purpose to ensure the validity and user guidance. Too many configurators are developed like for feature models commercial configurators [22] have been designed. Open source configurators [23,24,25]are also available. For one-size-fits-all kernel configurators, grows for the purpose of increasing performance [26].

For purpose of bidirectional broadcasting the different devices like internet and PSTN are used. So there should be an embedded operating system used. Linux is open source. So it modified the kernel of embedded Linux for achieving the goal [27]. In FPGA embedded system a Linux operating system is used. Different fault injection techniques used for checking the reliability of the system [28]. Linux is quite good for different microprocessors due to robustness because it is more scalable and reliable [29].

## 2.1.3 QNX:

QNX software system limited designated the QNX for real time operating systems. It have the functionality of processing the medical machines, telecommunication, and electronic devices. The requirements of these devices are higher reliability and safety because all are attach with our daily life, also having the features of scalability and flexibility. It is small and fast.

Medical devices like calculated the genetic methodology can some time distributed on multi processors due to large amount of data. It is also provide fully networked applications and provide higher performance in a real time environment. The draft standard 1003.4 which acts in real time environment is structured for distributed transparent environment. It is uses with microkernel processors in real time environment which consists on the services of UNIX and POSIX [30].

**Table 1:** *A Comparison table of QNX, WinCE, Linux* [31]

| Features/OS | QNX | Window CE | Linux |
|---|---|---|---|
| **Processor** | X86-based | 32-bit address based | ARM, Strong ARM, MIPS, Hitachi SH, x86-compatiable |
| **Memory** | Small kernel, 12 KB | For kernel and communication support needs 350 KB | For Configurable kernel needs 125-256 KB |
| **Architecture** | Used Micro-kernel | Hieratical based | Layering structure based |
| **Scheduling algorithm** | Priority based, same priority then FIFO, Round Robin | Priority based time slice algorithms / 8 priority levels | Priority and quantum based / 4 classes of threads |
| **Preemptive/ Non-preemptive** | Pre-emptive | Pre-emptive thread scheduling | both |
| **Interprocessor communication** | Massage passing | Massage passing | Original linux IPC mechanism |
| **Network support** | Low-Level Network | Internet protocols (FTP, HTTP servers) | All current internet protocols |

## 2.2 Domain specific OS:

This section describes the software infrastructure like operating system for specifically developed for one domain called the sensor networks.

### 2.2.1 Embedded Sensor Network:

Wireless technologies helped in making the low cost sensor devices [32] which have the low power consumption. Sensors devices like Ad-hoc network have thousands of small devices spread across a large area and communication made through the sensing ability of that device by passing massages to each other. Military also use the sensing devices. The sensing networks its tracks every moment of enemy in war. If damage all other network due to an emergency situation sensor ad-hoc network are easy and quickly to establish and maintain. It also helped in making the large embedded sensor network. Embedded computer having sensors or actuators, for interaction with the local environment, in inexpensive way [33]. WSN's (wireless sensor networks) have the constraints so its protocols requirement designed with new applications. For long life time of a network it's necessary to have the capabilities of low power consumption and device complexity. Balance between communication through a network is necessary requirement (signal processing should be in a manner). Many applications of WSN is enabled by IEEE 802.15.4 [34]. A light weight operating system Contiki divided into two parts core and loaded programs is used for Wireless sensor networks. It have the event-driven kernel and also have the dynamic unloading and loading of different services. it is preemptive in multithreads environment and implemented in C language[35]. Some micro sensor nodes are Mote/TinyOS [36],Meta cricket [37], MIT location aware cricket [38], Bluetooth based sensor nodes[39] which operate with the help of MANTIS [40] operating system have the ability to manage multithreading, aggregation and time sensitive tasks. For supporting the multi-hop networking, a reservation-based NanoRK [41] real-time operating system is used. Pixie [42] OS is also used in sensor networks for resource managing in different applications. LiteOS [43] have the ability of managing WSN's.

**Table 2:** *Comparision of OS's Used in Sensor Networks.*[44,45,46]

| Features/OS | NanoRK | MANTIS | TinyOS | LiteOS | Contiki | SOS |
|---|---|---|---|---|---|---|
| **Real-time support** | Yes | No | No | No | No | No |
| **Architecture** | Monolithic | Layered | Monolithic | Modular | Modular | Modular |
| **Communication protocol** | Depends on socket based | MAC and COMM | Two multi-hop protocols 1.TYMO 2.dissemination | File consisted | uIP, TCP/IP protocol, Rime and both versions of IP, 4 & 6 | Massage based |
| **Static/dynamic Memory management** | Static management with no protection of memory | Dynamic memory management and no protection | Static Memory protection and management | Dynamic Management and protection of memory | Dynamic memory management and no protection at address spaced | Dynamic Management and protection of memory |
| **Language support** | C | C | NesC | LiteC++ | C | C |
| **Scheduling algorithm** | Harmonized and Monotonic based algorithms | Priority based classes | First in first out | Round robin | Priority based inturuppts | - |
| **Resource sharing** | Mutexe and semaphores | semaphores | Through virtualization | Sharing can done by preemptive synchronization | Through Serial Access | - |
| **Modals of** | Thread driven | Thread driven | Event and thread | Thread and | Event and protothreads | Event driven |

## 3. Characterizations of embedded systems:

Embedded means that a system inside another system which work as a small and specific part of that system. Different embedded systems have different functionalities therefore there is no need to include all functionalities in an embedded system.

As we mentioned above the sensors are also embedded systems having a small processor, which having the ability of computation of physical environment like pressure, temperature etc. Embedded systems are divided into two categories which are given below.

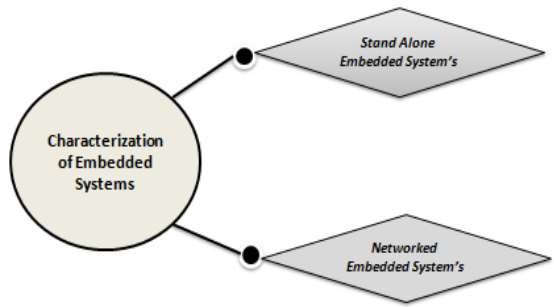

**Figure 2:** *Types of Embedded Systems*

### 3.1 Standalone embedded systems:

Users just give an input and the gets the desired output generated by the system are called stand-alone systems. Now these systems are said to be as system on chip and is the intellectual property. SOC's offers multimedia processors or DSP's application which needs the requirement of highly performance and embedded functionalities.

Some Standalone system includes the electronic consumers. CE divided into three types, Low-end devices are normally inexpensive cheap processors and System on Chip devices having the minimum memory like ROM (256 KB). Mid-range devices having memory of almost 1 to 2 MB, like digital camera etc. High-end consumer devices having the memory up to 32 MB , devices like smart-phones are included in it.

### 3.2 Networked embedded Systems:

There are many types of NES's some form these systems are network embedded system technology (NEST) , network systems for embedded computers (EmNETS ), Distributed real time embedded (DRE). Communications between functionally Distributed nodes which can be embedded are connected through the wireless/wired medium. it achieves the goal of interaction with environment by some actuators and sensing devices.

Paper [47] represents the general network embedded systems which includes distributed and real-time embedded systems, embedded systems and sensor systems. The reconfigurable, ubiquitous, and networked embedded systems called RUNES [48] develop a programing platform to for developing architecture of the software's used in NES's.

Embedded systems mostly make there computations on the devices which make for other purposes. It is not necessary that all devices are connected and usually not a mobile device. It having one server, and have wired connection.

Sensor Networks have a famous example of wireless sensor network. It gets a great deployment of various types of applications in industrious environment and in our daily life, including health and military applications. Sensor network[49] consist on the small , low power consumption machines which have the ability of concurrent execution of reactive programs that must operate with the constraints of space and low power. Different programming model can be supported by NesC complier which analyzes the overall program, also maintained the inline function and reliability. NesC and tinyOS used by many of sensor networks. The key challenge for wireless network embedded systems is power supplying. It can be done by solar energy [50] several factors given in the paper which should be understandable.

Distributed real-time and embedded systems are mostly time critical. In different conditions which are time and life critical systems should must have the accuracy like aircraft systems.[51]

A NIST Net[52] emulation tool is used for experiments on the different network protocols and performance of the network.

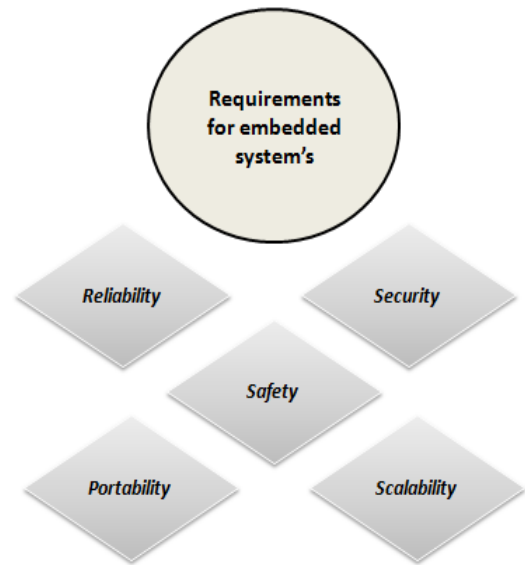

Figure 3: *Requirements in Embedded Software's*

## 4. Requirements of embedded systems:

There are two types of requirement in software engineering. Functional requirements define the function of a software component. Describe the set of inputs and generate the desired output. Non-Functional requirements are requirement which is used for judgment of operation of the system rather than the given behaviour of the systems also called qualities of system. There are some requirements of embedded systems are given below.[53]

**Reliability:** The ability of a system which regularly perform their specific task on demand and on perfect time. No faults occur called the system reliability**.** A study of software reliability shows that there can be tolerated six-fifteen bugs per thousand lines codes according to this LINUX has 2.5 million lines of code and window XP has double of this code so according to the software reliability study window XP kernel has almost 30000 bugs which make it unreliable and in secure. Micro kernels have the higher reliability but lower performance than monolithic kernels. May be it come back in near future due to reliability [54].

Embedded system should be highly reliable because fixing the fault in embedded system is highly challenging task. In the performance requirement of a system reliability is the main feature [55].

**Portability:** A system should have the portability means if the system is developed in one environment it should execute in other environment with very minor change. Portability and adaptability is an important feature of embedded system [56,57].

The internal quality of a system or system software is portability, which includes Adaptability, Installability, Co-existence, Replaceability and Compliance [58].

**Safety:** Some embedded systems are safety critical which should not harm the human life like aircraft system, or insulin pump system which is life critical. System safety is defined by engineers as risks [59].In a paper XtratuM [60] is designated for completing the requirement which are safety critical. Another CESAR [61] methodology provides a platform in safety critical application in different domains.

**Security:** In new electronic devices data security is very critical task includes computer, mobile phones etc. All the stages of an embedded system design from scratch to their development should be secure [62]. There is a large rang of embedded systems which have the issue of security for data manipulating, storing and sending etc. In networking security [63] protocols should be included for safe transmission of data.

Scheduling algorithms is failed to do the certain task in embedded systems because many of them ignores the security requirement. The given paper [64] applied the time constraints and security on the periodic task scheduling algorithms used in embedded systems.

Scalability, upgradability and fault tolerance are also requirements for embedded systems.

## 5. Conclusion:

As the above study of different software's which runs on embedded systems should have the requirements of safety and security. But many of them don't have the requirement and not support the real-time constraints and ability of fault tolerance. Research going own in these days, make the software's of embedded systems more reliable and include all the requirement of normal system. Network embedded systems are security critical for secure communication. The comparison of the sensor network OS's are given which shows the abilities and support that can be handled by the system. In these days, It is a challenging task that OS's used in network embedded systems should fulfils all the requirements like other systems.